\documentclass[%
 reprint,
 amsmath,amssymb,
 aps,
]{revtex4-1}

\usepackage{graphicx}
\usepackage{dcolumn}
\usepackage{bm}
\usepackage{amsmath,amssymb,bbm}

\usepackage{hyperref,float}

\usepackage[usenames,dvipsnames]{xcolor}
\newcommand{\lc}{\text{L}}
\newcommand{\agg}{\text{G}}
\newcommand{\ER}{Erd\H{o}s-R\'{e}nyi }

\begin{document}
\preprint{APS/123-QED}

\title{Trend-driven information cascades on random networks}

\author{Teruyoshi Kobayashi}
 \homepage{kobayashi@econ.kobe-u.ac.jp}
\affiliation{Graduate School of Economics, Kobe University, 2-1 Rokkodai, Nada, Kobe 657-8501, Japan} 
\date{\today}%
\vspace*{.5cm}

\begin{abstract}
 Threshold models of global cascades have been extensively used to model real-world collective behavior, such as the contagious spread of fads and the adoption of new technologies. A common property of those cascade models is that a vanishingly small seed fraction can spread to a finite fraction of an infinitely large network through local infections.
 In social and economic networks, however, individuals' behavior is often influenced not only by what their direct neighbors are doing, but also by what the majority of people are doing as a trend.  A trend affects individuals' behavior while individuals' behavior creates a trend. 
 To analyze such a complex interplay between local- and global-scale phenomena, I generalize the standard threshold model by introducing a new type of node, called \textit{global nodes} (or \textit{trend followers}), whose activation probability depends on a global-scale trend; specifically the percentage of activated nodes in the population. The model shows that global nodes play a role as accelerating cascades once a trend emerges while reducing the probability of a trend emerging. 
Global nodes thus either facilitate or inhibit cascades, suggesting that a moderate share of trend followers may maximize the average size of cascades. 
\begin{description}
\item[PACS numbers]
64.60.aq, 
89.65.-s. 
\end{description} 
\end{abstract}

\maketitle

\section{Introduction}

The mechanisms underlying various sorts of cascades on complex networks have been extensively studied over the past decade~\cite{Watts2002,Gleeson2007,Gleeson2008,Melnik2013,Nematzadeh2014,Liu2012,Payne2009,Payne2011,Centola2007,Watts2007}. The threshold model of cascades formalized by Watts~\cite{Watts2002}, among others, has been widely employed to describe not only information cascades on social networks~\cite{Watts2002,Watts2007}, but also default contagion in financial networks~\cite{GaiKapadia2010,Gai2011,Kobayashi2014,Brummitt2015} and cascades on multiplex networks~\cite{Lee2014,Yagan2012,Brummitt2012_PRER}. 

 A common assumption implicitly made in the threshold models is that a node's activation probability depends only on local information: the state of neighbors. 
 The point of the Watts model is that a vanishingly small seed fraction can spread to a finite fraction of an infinitely large network through local infections, suggesting that individuals' behavior at the local scale can add up to a global-scale outcome. 
 
 In real-world social and economic networks, nodes (individuals, firms, banks, etc.) normally collect information not only from their direct neighbors, but also from a wide variety of media sources. 
 Individual nodes can easily access global-scale information (poll results, average prices, average loan rates, etc.) and use it in their own decision-making (whom to vote for, whether to raise or maintain prices, whether or not to change loan rates, etc.). 
  This implies that a global-scale outcome affects individuals' behavior while individuals' behavior adds up to a global-scale outcome.
 Thus, there arises a complex interplay between local- and global-scale phenomena.
  
 To analyze such an interplay in a formal way, I generalize the Watts' threshold model by considering two types of nodes: \textit{local nodes} (or \textit{trendsetters}) and \textit{global nodes} (or \textit{trend followers}). The former are activated only by local neighbors as in the standard model, while {global nodes} are nodes whose states are dependent on a global-scale outcome. In the model, global nodes will be activated if a large enough fraction of the total number of nodes are activated. 
 
  The main contribution of this paper is to show that the introduction of global nodes significantly changes the mechanism of how a large-size cascade appears.
 There arises a saddle-node bifurcation in the presence of global nodes, and thereby a fold catastrophe may occur. This is intuitively because once the global nodes begin to be activated, those activated nodes further activates the local nodes that are otherwise inactive, leading to a drastic cascade. We call this type of drastic cascade the \textit{trend-driven cascade}. 
 
 On the other hand, the global nodes may also inhibit cascades, because they are never activated until the cascade size exceeds a certain threshold value.
 Thus, three phases may exist in the model: {Phase I}: no nodes are activated (except for seed nodes); {Phase II}: only local nodes may be activated; and {Phase III}: every node may be activated.
 It is shown that the boundary between {Phase II} and {Phase III} is well approximated by a root of a fixed-point equation, while the extended cascade condition proposed by Gleeson and Cahalane~\cite{Gleeson2007} can still accurately detect the boundary between Phase I and Phases II$+$III. 
 
 The analysis also reveals that the average size of cascades will be uniquely maximized near the boundary between Phase II and Phase III, where there are a medium fraction of global nodes in the population.
 These results suggest that the share of trend followers in the total population may be a key to understanding real-world collective behavior.

\section{A generalized threshold model with trend followers} 

\subsection{Two types of nodes}

 We consider an undirected and unweighted network that consists of $N$ number of nodes. The degree of each node is denoted by $k$. 
 Each node is initially ``inactive", except for the seed nodes that are originally ``active" by definition. The seed nodes are selected uniformly at random with probability $\rho_0$.
  The response function for the \textit{local nodes} (or \textit{trendsetters}) is 
 \begin{align}
  F^{\lc}(m,k) 
   &\equiv \begin{cases} 1 & \text{if } \frac{m}{k} > \theta_{\lc} \\ 0 & \text{otherwise} \end{cases} \label{eq:response_local}
 \end{align}
where $m$ denotes the number of neighbors that are already active and $\theta_\lc\in (\rho_0,1]$ is the threshold above which local nodes are activated. 

The response function for the \textit{global nodes} (or \textit{trend followers}) is given by
\begin{align}
  F^{\agg}(\rho) 
   &\equiv \begin{cases} 1 & \text{if } \rho > \theta_{\agg} \\ 0 & \text{otherwise} \end{cases} \label{eq:response_aggregate}
 \end{align}
where $\rho$ is the share of activated nodes in the population, and $\theta_\agg\in (\rho_0,1]$ is the threshold value. 
The essential difference between the two types of response functions is that $F^{\lc}$ is a function of $m/k$, which is the share of activated nodes in a \textit{vanishingly small fraction} of nodes
(i.e., $k/N$ approaches 0 as $N$ goes to infinity), whereas $F^{\agg}$ is a function of the share of activated nodes in a \textit{finite fraction} of nodes.
In this paper, we focus on a particular case in which the global nodes are activated if a large enough fraction of 
the total population exceeds a threshold. 
Note that the value of $\rho$ is global-scale information that is not identifiable from the states of local neighbors. The accessibility to global-scale information can be justified by assuming that each node has access to media, such as web sites, TV news, newspapers, etc.

 In numerical simulation, a global node is activated at step $t$ if the total share of activated nodes at step $t-1$ exceeds $\theta_\agg$. The local and global nodes are selected uniformly at random with probabilities $1-\mu$ and $\mu$.

\subsection{Treelike approximation for the average size of cascades}

 A treelike approximation for the conditional size of cascades, denoted by $\rho_\infty$, is given by a fixed point of the following recursion equations~\cite{Gleeson2007,Gleeson2008}
 \begin{align}
 \rho_{t+1} &= (1-\mu)\rho_{t+1}^{\lc} + \mu\rho_{t+1}^{\agg}, \label{eq:recursion_rhot}\\
 \rho_{t+1}^\lc &= \rho_0 + (1-\rho_0)\sum_{k=0}^{\infty}p_k\sum_{m=0}^{k}B_m^k(\tilde{\rho}_{t})F^{\lc}(m,k),\\
 \rho_{t+1}^{\agg} &= \rho_0+(1-\rho_0)F^{\agg}(\rho_{t}), \label{eq:recursion_rhot_agg}
 \end{align}
 where $p_k$ denotes the degree distribution and $B_m^k(\tilde{\rho}_t) \equiv \binom{k}{m} \tilde{\rho}_t^m (1-\tilde{\rho}_t)^{k-m}$. $\rho_{t+1}^{\lc}$ and $\rho_{t+1}^{\agg}$ are the fraction of active nodes at step $t+1$ among local and global nodes, respectively. Note that in Eq.(\ref{eq:recursion_rhot_agg}), $\rho_{t+1}^{\agg}$ is determined independently of the degree distribution. This is because global nodes are assumed to ignore local information coming directly from their neighbors.
 $\tilde{\rho}_t$ is obtained as a solution to the following recursion equations
 \begin{align}
  \tilde{\rho}_t &= (1-\mu)\tilde{\rho}_t^{\lc} + \mu\tilde{\rho}_t^{\agg}, \\
 \tilde{\rho}_{t}^\lc &= \rho_0 + (1-\rho_0)\sum_{k=1}^{\infty}\frac{kp_k}{z}\sum_{m=0}^{k-1}B_m^{k-1}(\tilde{\rho}_{t-1})F^{\lc}(m,k),\\
 \tilde{\rho}_{t}^{\agg} &= \rho_t^{\agg}, \label{eq:recursion_rhotildet_agg}
 \end{align}
where $z$ is the mean degree. Thus, six variables  $\rho_{t+1},\rho_{t+1}^{\lc},\rho_{t+1}^{\agg},\tilde{\rho}_t,\tilde{\rho}_t^{\lc}$ and $\tilde{\rho}_t^{\agg}$ are determined by six equations (\ref{eq:recursion_rhot})-(\ref{eq:recursion_rhotildet_agg}), taking $\tilde{\rho}_0 = {\rho}_0^\lc = {\rho}_0^\agg = \rho_0$ as given. 
Notice that $\rho_\infty$ may not be interpreted as the average size of cascades because the probability of a global cascade occurring is not necessarily equal to 1.
 One should interpret $\rho_\infty$ as the size of the cascade, or equivalently as the probability of a randomly selected node being activated, \textit{conditional on} the event that at least one seed node belongs to the extended vulnerable cluster~\cite{Gleeson2008}.
 

 As discussed by Gleeson~\cite{Gleeson2008} for the standard case of $\mu=0$, $\rho_\infty$ becomes an accurate approximation for the average size of cascades if the initial seed fraction $\rho_0$ is not vanishingly small, where the frequency of global cascades will be sufficiently close to 1. If instead $\rho_0$ is vanishingly small, then we need to multiply $\rho_\infty$ by the probability that a global cascade occurs. This is because there arises a non-negligible possibility that all the seed nodes may be located outside the \textit{extended vulnerable cluster}~\cite{Watts2002,Gleeson2008}, in which case the cascade probability will be less than 1. In the presence of global nodes, the size of the extended vulnerable cluster becomes smaller than that of the conventional one since global nodes are ``immune" to local infection. The frequency of global 
 cascades can be significantly lower than 1 even if $\rho_0$ is not vanishingly small. 
 
 The average size of the \textit{modified extended vulnerable cluster} $\tilde{S}_e$ is given as
 \begin{align}\label{eq:extended_cluster}
  \tilde{S}_e &= \sum_{k=0}^{\infty}p_k[1-(1-\tilde{q}_\infty)^k], \\
   \tilde{q}_{t} &= (1-\mu)\sum_{k=1}^{\infty}\frac{kp_k}{z}[1-(1-\tilde{q}_{t-1})^{k-1}]F^{\lc}(1,k).
 \end{align}
 Note that a global cascade occurs if and only if at least one seed node belongs to the modified extended vulnerable cluster.
The (unconditional) average size of cascades, denoted by ${S}$, is approximated as
  \begin{align}
  {S} \approx \rho_\infty [1-(1-\tilde{S}_e)^{\lfloor \rho_0 N\rfloor}]. \label{eq:S}
  \end{align}
The terms in the square brackets capture the (approximated) probability that at least one seed node belongs to the modified extended vulnerable cluster.

 \section{Analysis}
 \subsection{Cascade conditions}
 
 The parameter space within which a global cascade may occur is called the \textit{cascade region}, and the condition that determines whether or not a global cascade may occur is called the \textit{cascade condition}.
 The \textit{first-order cascade condition} is a cascade condition with first-order accuracy while the \textit{extended cascade condition} guarantees second-order accuracy~\cite{Gleeson2007}. 
 
 Appendix~\ref{sec:cascade_condition} shows the derivation of the first-order and extended cascade conditions. Note that although our model is summarized as a system of two recursion equations, Eqs.(\ref{eq:appendix_rhotilde})-(\ref{eq:appendix_rhot}), the only equation relevant to the cascade conditions is
\begin{align}
 \tilde{\rho}_{t} &= (1-\mu)\big[\rho_0 + (1-\rho_0)\sum_{k=1}^{\infty}\frac{kp_k}{z}\nonumber \\ & \times \sum_{m=0}^{k-1}B_m^{k-1}(\tilde{\rho}_{t-1}) F^{\lc}(m,k)\big].\label{eq:cascadecondition_rhotilde}
\end{align}
In fact, this recursion equation is the same as that obtained in a hypothetical model in which there were no global agents and $100\mu\%$ of local nodes were randomly removed. The cascade region that appears in the present model coincides with that of the hypothetical model, which implies that it is only the local nodes that can initiate cascades.
This irrelevancy of the global nodes stems from the fact that the global nodes are totally unresponsive in the initial stage of contagion under the natural assumption of $\lim_{\rho_{t-1}\downarrow 0}\frac{d}{d\rho_{t-1}}F^\agg(\rho_{t-1})=0$.  The cascade condition in the presence of global nodes is the one that indicates the parameter space within which a finite fraction of the local nodes can be activated.

\begin{figure}
\includegraphics[width=1.0\columnwidth,clip]{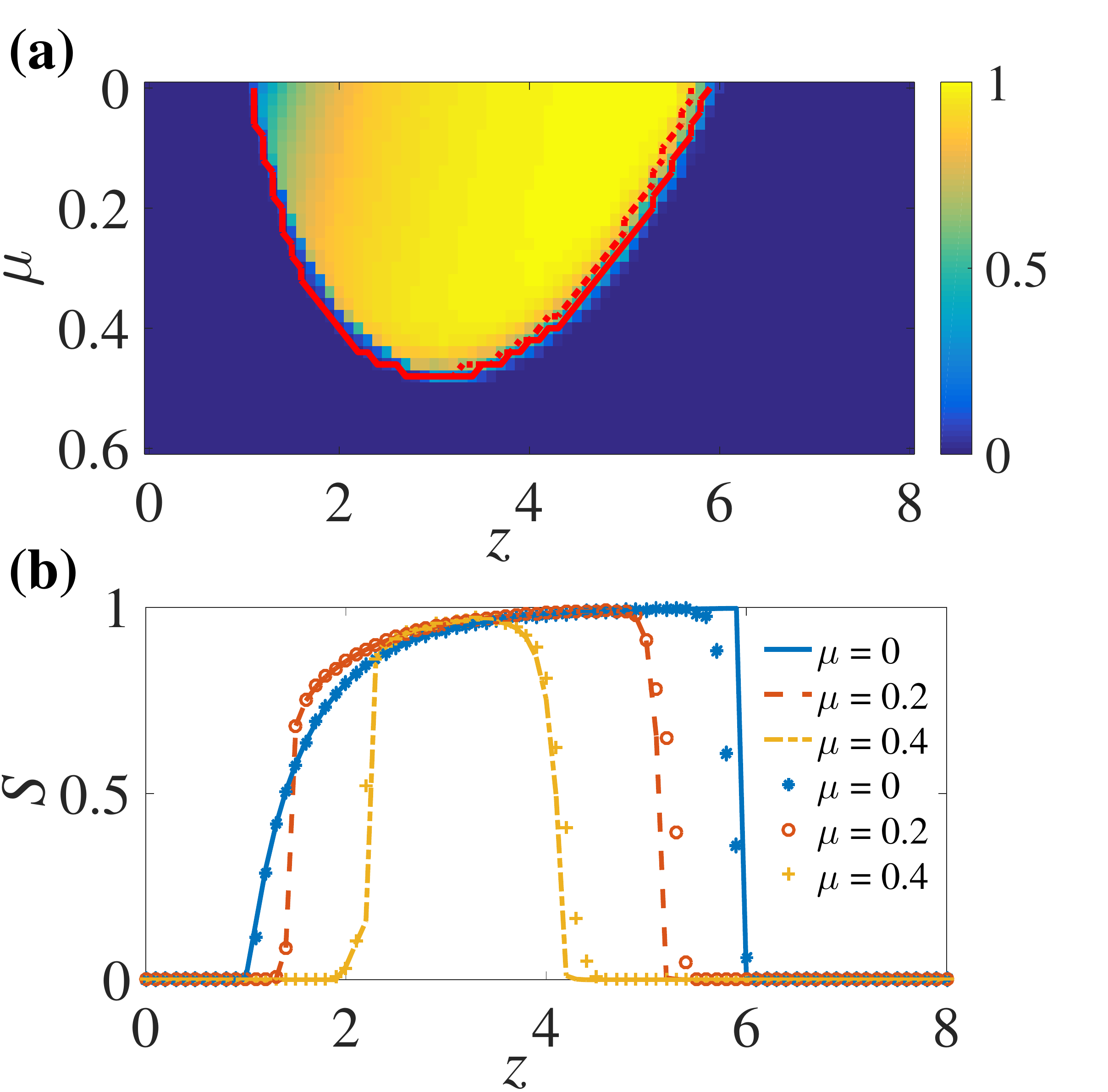}
\caption{(Color online) Percentage of global nodes in the population, $\mu$, and average size of cascades on \ER networks.
(a) Simulated average cascade size (color bar) and the cascade conditions. Dashed and solid lines denote the first-order and the extended cascade conditions, respectively. 
(b) Lines denote the theoretical average size of cascades, {$S$}, and markers plot simulated values.
$N=10^5$, $\rho_0=10/N$ and $\theta_\lc = \theta_\agg = .18$, averaged over $10^3$ random perturbations.}
\label{fig:fixedthreshold_baseline}
\end{figure}

 Fig.~\ref{fig:fixedthreshold_baseline} demonstrates that our analytical approximation matches reasonably well with simulations on \ER random networks. Fig.~\ref{fig:fixedthreshold_baseline}(a) compares the simulated value of $S$ and the cascade conditions. It is known that the first-order cascade condition (dashed line) is not very accurate for a finite seed size~\cite{Gleeson2007,Gleeson2008}, but the extended cascade condition (solid line) still correctly predicts the upper as well as the lower phase transition. 
 If we set $\rho_0=\mu=0$ in the first-order cascade condition, then it predicts the cascade region shown in the original Watts model~\cite{Watts2002}.

 \subsection{Cascade region for trend-driven cascades}

\begin{figure}
\includegraphics*[width=1.0\columnwidth,clip]{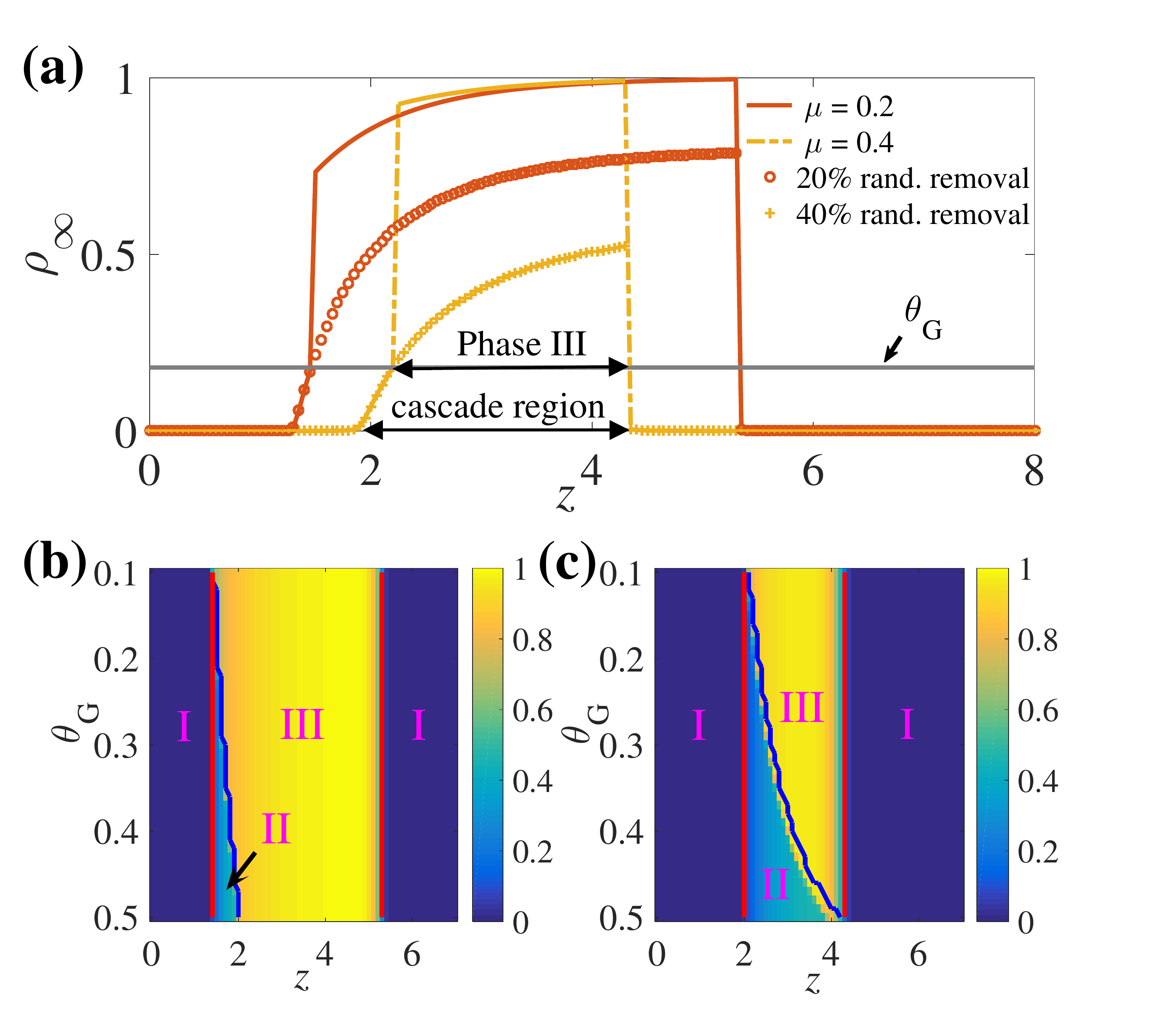}
\caption{(Color online) Trend-driven cascades. (a) Lines denote $\rho_\infty$ obtained from Eqs.~\eqref{eq:recursion_rhot}--\eqref{eq:recursion_rhotildet_agg}, while markers plot the corresponding value obtained in a model without global nodes but $100\mu\%$ of nodes being removed at random. (b) and (c) Colors indicate the simulated $S$ and red solid line denotes the extended cascade condition. Blue dash-dotted line denotes an approximated lower boundary $\tilde{z}$ above which a trend-driven cascade occurs.
 Roman numerals denote the phase of the region.  $N=10^5$, $\rho_0=10/N$, $\theta_\lc=.18$, $\theta_\agg = .18$ [in panel (a)], averaged over $10^3$ random perturbations.}
\label{fig:fixedthreshold_newscascade}
\end{figure}

 The commonality in the cascade regions between the present model and the hypothetical model with random node removal is illustrated in Fig.\ref{fig:fixedthreshold_newscascade}(a). The figure shows that the conditional cascade size, $\rho_\infty$, is identical in the two cases for $\rho_{\infty} < \theta_\agg$. For $\rho_\infty > \theta_{\agg}$, $\rho_\infty$ in the current model becomes larger than that under the hypothetical model because the ``removed" nodes will be activated. Note that the size of $\rho_\infty$ in the presence of global nodes increases discontinuously at $\hat{z}$, where $\hat{z}\equiv\inf\{z: \rho_\infty > \theta_\agg\}$.
 This indicates that the parameter space can be decomposed into three parts: Phases I, II and III [Figs.\ref{fig:fixedthreshold_newscascade}(b) and (c)]. Phase I is the region in which no nodes are activated (except for the seed nodes). Phase II is the cascade region within which only local nodes can be activated, while every node may be activated in Phase III. 
 
 For \ER random graphs, an approximated value of $\hat{z}$, denoted by $\tilde{z}$, is obtained as a solution to the following root-finding problem obtained from Eq.~\eqref{eq:cascadecondition_rhotilde}
  \begin{align}
   & (1-\mu)\big[\rho_0 + (1-\rho_0)\sum_{k=1}^\infty\frac{\tilde{z}^{k-1}e^{-\tilde{z}}}{(k-1)!} \nonumber \\ &\times \sum_{m=0}^{k-1}B_m^{k-1}(\theta_\agg) F^{\lc}(m,k)\big] - \theta_\agg =0   , 
  \label{eq:critical_cascade}
  \end{align}
and subject to $\tilde{z}\in [\underline{z},\bar{z}]$, where $\underline{z}$ and $\bar{z}$ are the lower and upper boundaries of the extended cascade region, respectively. The upper boundary of Phase III is given by $\bar{z}$. Note that because $\tilde{\rho}_\infty$ and $\rho_\infty$ can generally differ, $\tilde{z}$ obtained by solving~\eqref{eq:critical_cascade} is an approximated, rather than the correct, lower boundary of Phase III. Nevertheless, Figs.\ref{fig:fixedthreshold_newscascade}(b) and (c) demonstrate that the so-defined $\tilde{z}$ nicely matches the numerical lower edge.

\subsection{Bifurcation analysis and tricritical-point scaling}

\begin{figure}
\includegraphics[width=1\columnwidth,clip]{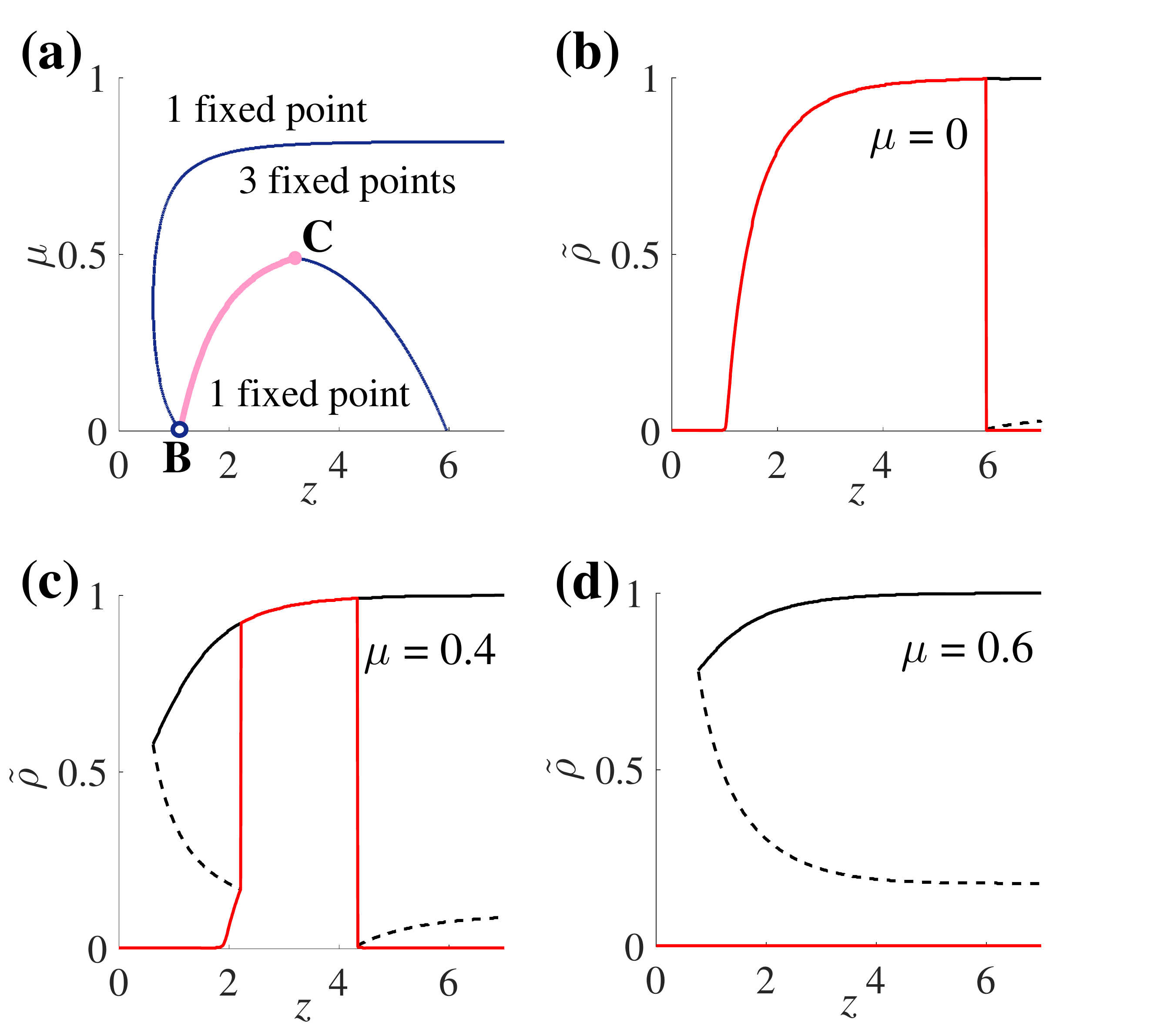}
\caption{(Color online) Bifurcation diagram of the roots $\tilde\rho$ of fixed-point equations $(\tilde{\rho},\rho)-g(\tilde{\rho}, \rho )=0$ [eqs.~\eqref{eq:appendix_rhotilde}--\eqref{eq:appendix_rhot}]. 
(a) Stability diagram that distinguishes the 
parameter space $(z,\mu )$ in terms of the number of fixed points. The first-order line $\mathbf{BC}$ highlighted in pale red denotes a set of combination $(z,\mu )$ on which a trend-driven cascade (i.e., a {fold} catastrophe) occurs. 
 Panels (b)--(d) Bifurcation diagram for alternative values of $\mu$.
 Black solid and dashed lines denote stable and unstable fixed points, respectively. Red line is the realized stable root for cascades starting from the initial seed fraction $\rho_0=10^{-4}$.  
$\theta_\lc = \theta_\agg = .18$.} 
 \label{fig:bifurcation_diagram}
 \end{figure}

The presence of discontinuous transitions can also be confirmed by the bifurcation diagram illustrated in Fig.~\ref{fig:bifurcation_diagram}.
{As pointed out by Gleeson and Cahalane~\cite{Gleeson2007}, the high-$z$ phase transition (i.e., the second phase transition when $\mu =0$ and the third phase transition when $\mu > 0$) becomes discontinuous due to a saddle-node bifurcation. Notice that in this model $\mu$ must be sufficiently small for the phase transitions to occur (approximately $\mu < 0.5$ under the baseline parameters).}

{Our particular interest is on fold catastrophes that occur when the parameter combination $(z,\mu)$ is on the bifurcation curve $\mathbf{BC}$ highlighted in pale red, which is called the \textit{first-order line} (Point $\mathbf{C}$ is called a cusp point~\cite{Strogatz2014}). 
 Given $\mu = 0.4$, for instance [Fig.~\ref{fig:bifurcation_diagram}(c)], the first phase transition occurring at $z\approx 1.9$ (from Phase I to Phase II) is smooth, but the second transition (from Phase II to Phase III) at $z= 2.3 (\approx\tilde{z} )$ is discontinuous. 
The introduction of global nodes distinguishes the first smooth transition from the second discontinuous one however small the value of $\mu (>0)$ is.
Note that the classical cascade region is not identical to the region labeled as ``1 fixed point" in Fig.~\ref{fig:bifurcation_diagram}(a), because there can be three fixed points in phase II.}
 Recently, Lee et al.~\cite{Lee2014} also show that a saddle-node bifurcation emerges in a multiplex network when there is heterogeneity in the response functions. 

\begin{figure}
\includegraphics[width=1\columnwidth,clip]{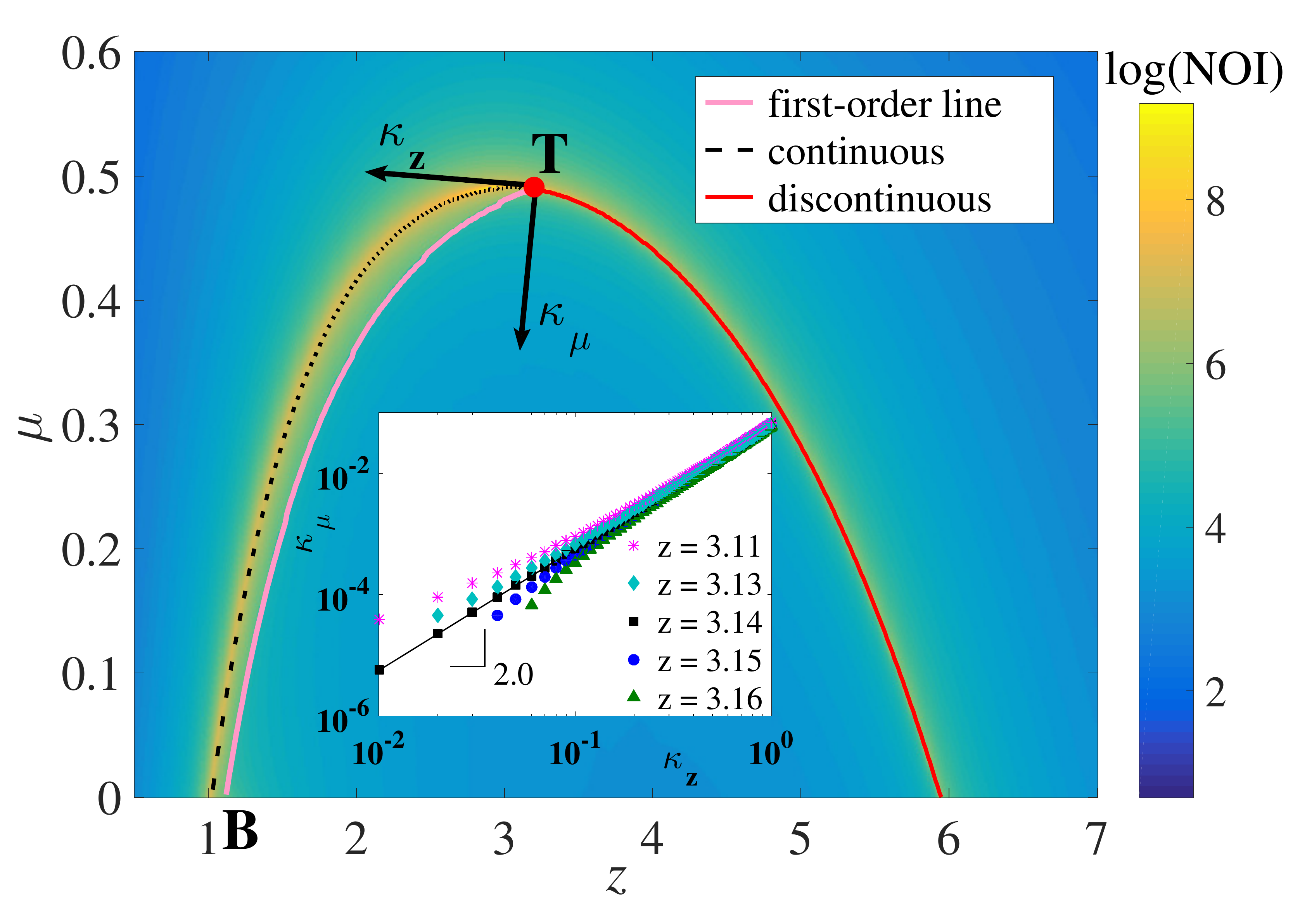}
\caption{(Color online)
Scaling at the tricritical point $\mathbf{T}$. NOI denotes the number of iterations needed to ensure that the successive iterates in the recursion equations~\eqref{eq:appendix_rhotilde}-\eqref{eq:appendix_rhot} differ by less than $10^{-10}$. The critical line (or the cascade region) is identified by the local maxima of the NOI.
The critical line and the first-order line pass through the tricritical point $\mathbf{T}=(3.14,0.49)$, where the global maximum of the NOI is attained. 
(Inset) Scaling of the critical line in a new coordinate system $(\kappa_z,\kappa_\mu)$ centered at a point on the critical line. The log-log plot demonstrates that the critical line in the new coordinate system centered at $\mathbf{T}$ (black squares) fits a straight line with slope 2.0 near the origin. To obtain the new coordinate system, the numerical critical line is smoothed by a 7th-order polynomial function. $\theta_\lc = \theta_\agg = .18$.
}
 \label{fig:bifurcation_scaling}
\end{figure}

Fig.~\ref{fig:bifurcation_scaling} illustrates the boundary of the classical cascade region, called the \textit{critical line}, detected by finding the local maxima of the number of iterations (NOI) in the recursion equations~\eqref{eq:appendix_rhotilde}-\eqref{eq:appendix_rhot}. It demonstrates that as $z$ increases, the NOI {changes} drastically when crossing the critical line and the first-order line. Both lines pass through the tricritical point $\mathbf{T}\equiv (z_t,\mu_t)=(3.14,0.49)$ (subscript $t$ stands for ``tricritical"), {where phases I, II and III terminates. The NOI attains the global maximum at $\mathbf{T}$~\cite{Lee2014}.} The tricritical point $\mathbf{T}$ and the cusp point $\mathbf{C}$ in Fig.~\ref{fig:bifurcation_diagram}(a) generally differ {(recall that the cascade region is not distinguished by the number of roots $\tilde\rho$)}, but in the present case they are very close.

 The inset figure shows the scaling of the critical line 
in the scaling fields $(\kappa_z,\kappa_\mu)$~\cite{Riedel1972,Araujo2011,Cellai2011,Lee2014}, where $\kappa_z$ is tangent to the critical line at $\mathbf{T}$ and $\kappa_\mu$ is perpendicular to it. 
A new coordinate system is given by the transformation
\begin{align}
\begin{pmatrix}
 \kappa_z \\ \kappa_\mu
\end{pmatrix}
= \begin{bmatrix}
 \cos\tilde{\theta} & \sin\tilde{\theta} \\
 -\sin\tilde{\theta} & \cos\tilde{\theta}
\end{bmatrix}
\begin{pmatrix}
 z-z_t \\ \mu-\mu_t
\end{pmatrix},
\end{align}
where $\tilde{\theta}=\pi + \theta$, and $\theta\; (<0)$ is the angle of $\kappa_z$ with the $z$-axis.
In the new coordinate system, the critical line near the origin is described by
\begin{align}
 \kappa_\mu \sim \kappa_z^{1/\phi_t},\label{eq:scaling}
\end{align}
where $\phi_t$ is called the crossover tricritical exponent. In the baseline case we have $1/\phi_t \approx 2.0$. The log-log plot demonstrates that at a point slightly deviated from $\mathbf{T}$, the scaling does not follow Eq.~\eqref{eq:scaling}.

\section{Effect of trend-following behavior on the average size of cascades}
\subsection{Two opposite roles of trend followers}

\begin{figure}
\includegraphics[width=1\columnwidth,clip]{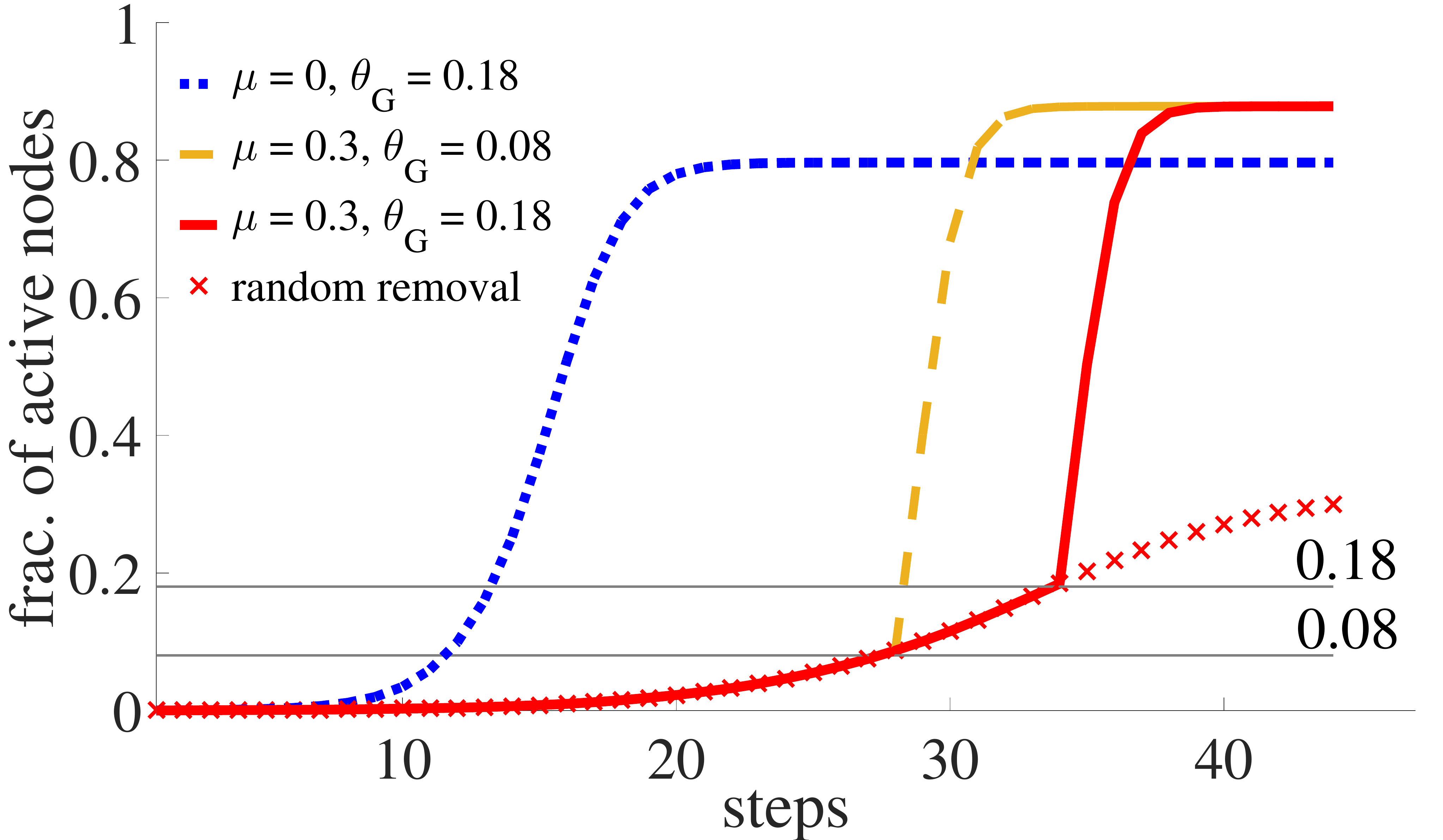}
\caption{(Color online) Cumulative fraction of activated nodes in a contagion simulation. A trend-driven cascade occurs when the cumulative fraction exceeds $\theta_\agg$. The path labeled as ``random removal" (red crosses) corresponds to the case in which $\mu =0$ but 30\% of the local nodes are randomly removed. 
$z=2$, $N=10^5$, $\rho_0=10/N$ and $\theta_\lc = .18$.}
\label{fig:fixedthreshold_path}
\end{figure}

 Fig.~\ref{fig:fixedthreshold_path} illustrates an example path of the cumulative fraction of activated nodes. When $z \in (\hat{z},\bar{z})$, the number of activated nodes in the steady state can be larger than that under $\mu=0$. 
 The figure also shows that the speed of convergence is decelerated by the introduction of global nodes that are never activated in the early stage of contagion.     

 It follows from the above results that the effect of introducing global nodes is twofold. First, cascades are inhibited by the presence of global nodes.
 The cascade region tends to shrink as the share of global nodes $\mu$ increases, and the speed of contagion decelerates accordingly. Global nodes play a role as ``immunized patients" that prevent infectious diseases from spreading widely.
 Fig.\ref{fig:fixedthreshold_cascadecondition} in Appendix~\ref{sec:cascade_condition} illustrates how the cascade region shrinks with $\mu$.

Second, the previous figures also demonstrate that the size of cascades may increase with $\mu$, provided that $z\in (\hat{z},\bar{z})$. 
An intuition is that once the total fraction of activated nodes exceeds $\theta_\agg$, all the global nodes are activated regardless of their degree, meaning that a certain proportion of nodes outside the largest connected component (LCC) can be activated. Recall that the size of global cascades is usually constrained by the size of the LCC when $\mu=0$~\cite{Watts2002,Gleeson2008}.  

In total, whether or not the introduction of global nodes increases the cascade size depends on the relative strength between the negative and positive effects.

\subsection{Optimal value of $\mu$}
\subsubsection{Maximizing the average cascade size}

\begin{figure}
\includegraphics[width=1\columnwidth,clip]{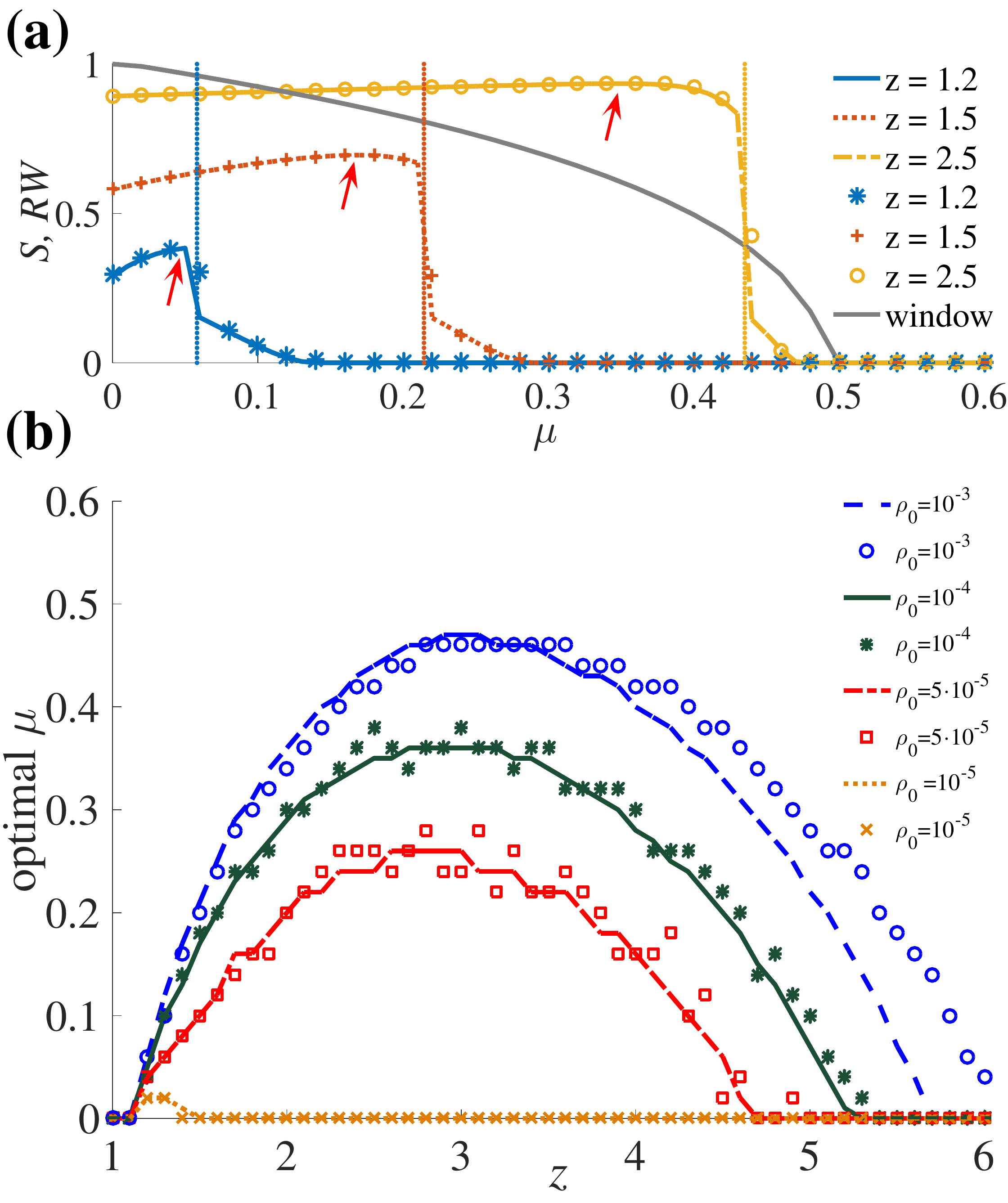}
\caption{(Color online) Optimal value of $\mu$. (a) Effect of varying $\mu$ on the average size of cascades. {Gray line denotes the relative size of the cascade window $RW$, defined as $RW(\mu)\equiv (\bar{z}-\underline{z})|_{\mu >0}/(\bar{z}-\underline{z})|_{\mu =0}$.} 
Vertical dotted lines denote $\bar\mu$ calculated by~\eqref{eq:muhat}. Other lines and markers show the theoretical and simulated values of $S$, respectively. The maximum of $S$ is pointed by arrow. (b) Optimal value of $\mu$ under alternative seed sizes. 
Lines and markers denote theoretical and numerical results, respectively. $N=10^5$, $\theta_\lc = \theta_\agg = .18$ and $\rho_0=10^{-4}$ [in panel (a)], averaged over $10^3$ random perturbations.}
\label{fig:fixedthreshold_optimalmu}
\end{figure}

Figure~\ref{fig:fixedthreshold_optimalmu}(a) illustrates how the average size of cascades $S$ varies with $\mu$, taking $z$ as given. It shows that ${S}$ initially increases due to the positive effect described above, but after passing a critical value, the negative effect comes to dominate the positive one and ${S}$ falls drastically {(Gray line shows the relative size of the cascade window, defined as $RW(m)\equiv (\bar{z}-\underline{z})|_{\mu >0}/(\bar{z}-\underline{z})|_{\mu =0}$)}. Thus, there exists a unique maximizer of $S$, which we call the ``optimal" value of $\mu$, denoted by $\mu^{*}$. 
The case of $\mu^* >0$ would be of interest because that is the case in which the presence of trend followers facilitates cascades.
A similar discussion on the maximization of information cascades is given by Nematzadeh et al.~\cite{Nematzadeh2014}. 

 Although an analytical expression for $\mu^*$ is difficult to obtain, we know that $\mu^*$ is within the parameter space that attains Phase III. From Eq.(\ref{eq:cascadecondition_rhotilde}), an approximated upper bound of $\mu^*$, denoted as $\bar{\mu}$, is given as 
 \begin{align}\label{eq:muhat}
    \bar{\mu} &= \max{\{0,\hat{\mu}\}}, \notag \\
    \text{where } \;\;
      \hat{\mu} &= 1-\theta_\agg\big[\rho_0 + (1-\rho_0)\sum_{k=1}^\infty\frac{kp_k}{z} \nonumber \\ &\times \sum_{m=0}^{k-1}B_m^{k-1}(\theta_\agg) F^{\lc}(m,k)\big]^{-1}  ,
 \end{align}
 taking the parameters other than $\mu$ as given. The dotted vertical lines in Fig.\ref{fig:fixedthreshold_optimalmu}(a) plot $\bar{\mu}$. 
  Another interpretation of $\bar{\mu}$ is that it indicates the minimum fraction of global nodes that is required to eliminate the possibility of trend-driven cascades. A further increase of ${\mu}$ from $\bar{\mu}$ will prevent any global cascades.  

Figure~\ref{fig:fixedthreshold_optimalmu}(b) compares the theoretical and simulated values of $\mu^*$. The theoretical value of $\mu^*$ is computed by a greedy search. It shows that the accuracy of theoretical $\mu^*$ gets worse as $z$ increases. This is because the larger the mean degree, the lesser the benefit of increasing $\mu$, which would undermine the accuracy of the greedy search in the presence of a finite-size problem. The average size of cascades is almost unchanged for $z>3$ since most of the nodes belong to the LCC. 

\subsubsection{Effect of varying the seed size}

 Fig.\ref{fig:fixedthreshold_optimalmu}(b) also states that $\rho_0$ and $\mu^{*}$ have a positive relationship.
 To understand this, recall that the increment of $S$ can be decomposed into two factors from the relation $S = \rho_\infty \Psi$, where $\Psi\equiv [1-(1-\tilde{S}_e)^{\lfloor \rho_0 N\rfloor}]$. 
A rise in $S$ stems from an increase either in $\rho_\infty$ or in $\Psi$ or in both. However, for a given $\mu$, the size of $\rho_\infty$ is determined solely by the size of the LCC independently of $\rho_0$ if $\rho_0 \ll 1$. The only factor that the initial seed size can affect is therefore $\Psi$, the (approximated) frequency of cascades. 

\begin{figure}
\includegraphics[width=1.0\columnwidth,clip]{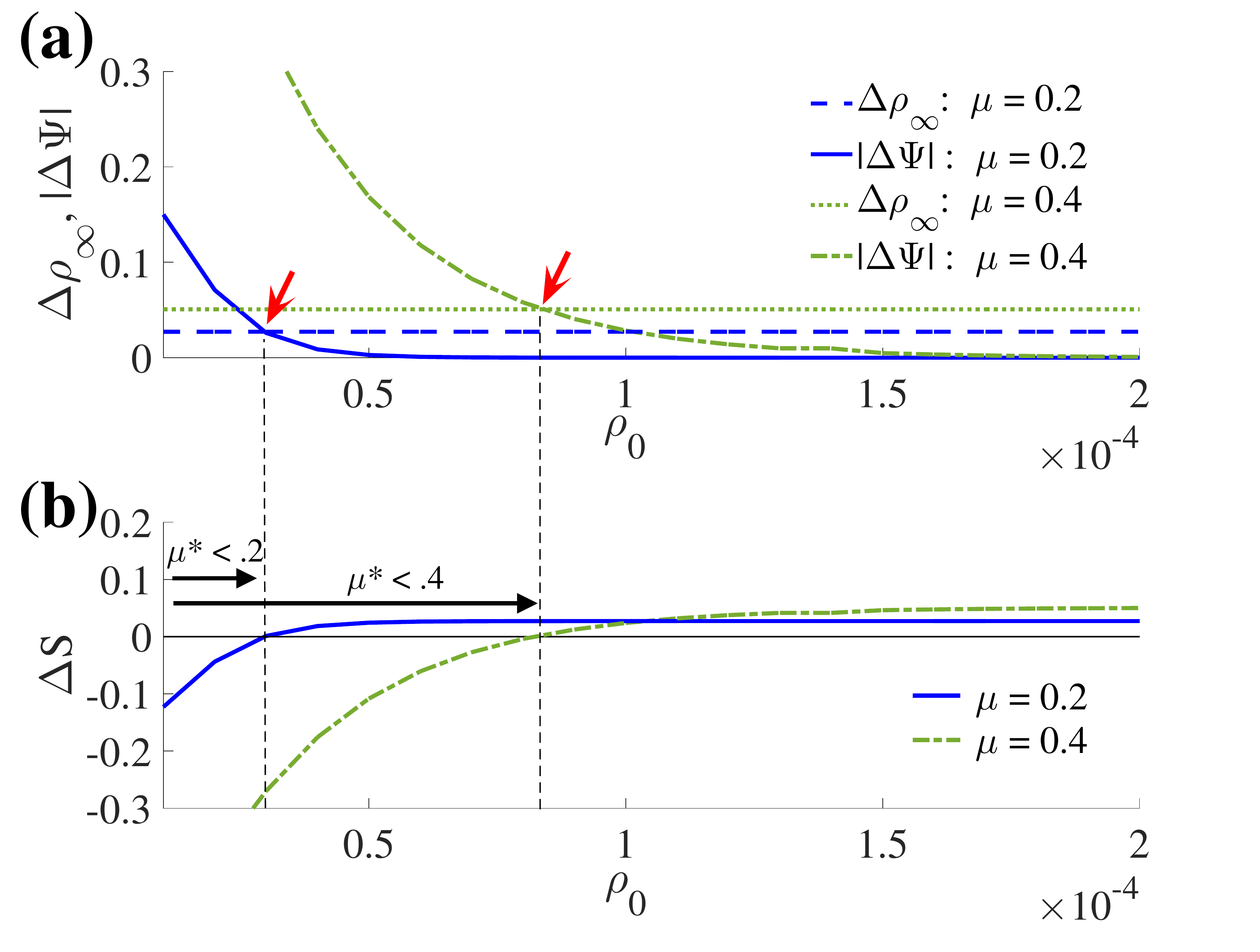}
\caption{(Color online) Decomposition of $\Delta{S}$. (a) Changes in $\rho_\infty$ and $\Psi$ caused by an increase in $\mu$. Arrows indicate the points above which a rise in $\rho_\infty$ dominates the reduction in $\Psi$, leading $\Delta{S}$ to be positive. (b) Change in the average size of cascades. 
$\theta_\lc = \theta_\agg = .18$, $z=2.5$.}
\label{fig:fixedthreshold_optimalmu_decomp}
\end{figure}

 Fig.~\ref{fig:fixedthreshold_optimalmu_decomp} illustrates the decomposition of an increment of $S$, defined as $\Delta{S}\equiv S|_{\mu=c}-S|_{\mu=0}$ for $c\in [0,1]$. In the figure, we see that $\Delta{\rho_\infty}\equiv \rho_\infty|_{\mu=c}-\rho_\infty|_{\mu=0}$ is independent of $\rho_0$, while $\Delta\Psi \equiv \Psi|_{\mu=c}-\Psi|_{\mu=0}$ $(<0)$ increases with $\rho_0$ because the larger the initial seed fraction, the higher the probability of a global cascade occurring.
  Suppose that $\mu$ has increased from $0.2$ to $0.4$. This largely lowers $\Psi$ by shrinking the size of the extended vulnerable cluster $\tilde{S}_e$, while increasing $\rho_\infty$ by activating more nodes located outside the LCC. The former effect dominates the latter for a small $\rho_0$, but if $\rho_0$ is large enough, then the increase in $\rho_\infty$ would compensate for the reduction in $\Psi$, which could make $\mu=0.4$ optimal.

\section{Conclusion and discussion} 
 
 In this work, I investigated an interplay between local- and global-scale contagion by generalizing the standard threshold model.
 In real-world social and economic networks, it is quite common that nodes respond to certain information in diverse ways.  For example, a retail firm may raise prices if some of its wholesalers (i.e., neighbors) increase their prices while others may set prices with reference to the rate of inflation (i.e., a global-scale trend).
  There is thus response heterogeneity among individuals, a topic also discussed in Lee et al.~\cite{Lee2014}. Interestingly, although Lee et al. introduced response heterogeneity in a multiplex version of the threshold model, they show a property similar to one suggested in the present work; a global cascade may appear \textit{abruptly yet slowly}.
 {In the context of multiplex networks, the current model could be expressed as a 2-layer multiplex model, where local and global nodes are located in different layers. The ``global layer" then forms a complete network where every node is connected to every other nodes, while the degree distribution in the ``local layer" is given by $p_k$. In that environment, the results of the current model will be replicated if it is assumed that fraction $\mu$ of the total nodes are activated only in the global layer and $1-\mu$ of nodes are activated only in the local layer. However, the relationship with the model of Lee et al.~\cite{Lee2014} is not straightforward because they consider two alternative response functions to capture the ``AND" and ``OR" strategies~\cite{Lee2014}.}         
  
Global cascade is not a phenomenon that appears in every social and economic network on a regular basis, suggesting that many real-world networks are stable.
According to the simple model presented here, the observed stability of actual networks may be attributed to the presence of trend followers who are activated only when a large enough fraction of the total population are activated. {The present model offers a possible explanation of why cascades are not very frequent in real-world networks as well as why large cascades may appear abruptly.} 
I hope this work will contribute to a better understanding of the interplay between local and global phenomena that could play a crucial role in the mechanism of collective behavior.

\section*{Acknowledgements}
Financial supports from the Japan Society for the Promotion of Science KAKENHI 25780203, 15H01948 and 15H05729 are gratefully acknowledged.

\appendix

\section{Cascade condition}
\label{sec:cascade_condition}

\begin{figure}
\includegraphics[width=1.0\columnwidth,clip]{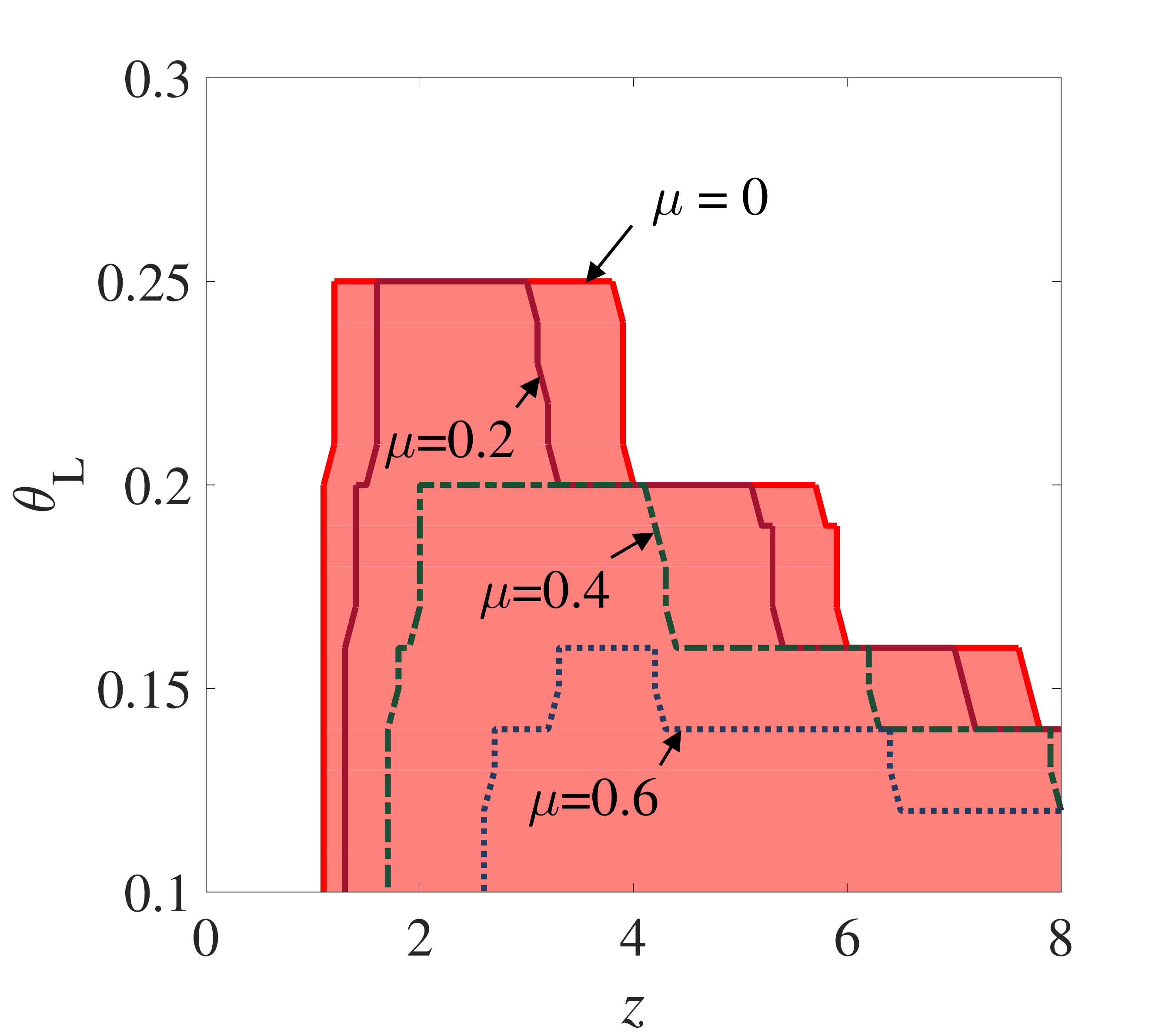}
\caption{(Color online) Extended cascade conditions under alternative values of $\mu$. Cascade region in the absence of global nodes ($\mu=0$) is painted in red.  
The dark-red solid ($\mu =.2$), green dash-dotted ($\mu=.4$) and blue dotted ($\mu=.6$) lines respectively indicate the outer boundary of the corresponding cascade region.
$\theta_\agg = .18$, $\rho_0=10^{-4}$.}
\label{fig:fixedthreshold_cascadecondition}
\end{figure}

 To obtain the cascade condition, we reorganize 
 Eqs.(\ref{eq:recursion_rhot}) - (\ref{eq:recursion_rhotildet_agg}) as the following two recursion equations that involve two variables, $\tilde{\rho}_t$ and $\rho_t$:
\begin{align}
 \tilde{\rho}_{t} &= (1-\mu)\big[\rho_0 + (1-\rho_0)\sum_{k=1}^{\infty}\frac{kp_k}{z} \sum_{m=0}^{k-1}B_m^{k-1}(\tilde{\rho}_{t-1})\notag \\ & \times F^{\lc}(m,k)\big]
+ \mu\big[ \rho_0+(1-\rho_0)F^{\agg}(\rho_{t-1})\big],\label{eq:appendix_rhotilde}\\
  \rho_{t} &= (1-\mu)\big[\rho_0 + (1-\rho_0)\sum_{k=0}^{\infty}p_k \sum_{m=0}^{k}B_m^{k}(\tilde{\rho}_{t-1})\notag \\ & \times F^{\lc}(m,k)\big]
 + \mu \big[ \rho_0+(1-\rho_0)F^{\agg}(\rho_{t-1})\big] ,\label{eq:appendix_rhot}
 \end{align}
  The first-order cascade condition is that the Jacobian matrix of function $g$, defined as $(\tilde{\rho}_t, \rho_t) = g(\tilde{\rho}_{t-1}, \rho_{t-1})$, has the largest eigenvalue larger than $1$ at $(0,0)$~\cite{Gleeson2007}. 
 However, since $\lim_{\rho_{t-1}\downarrow 0}\frac{d}{d\rho_{t-1}}{F^\agg}(\rho_{t-1})=0$, the Jacobian has rank one, meaning that the trace is the only nonzero eigenvalue. Therefore, the first term in Eq.\eqref{eq:appendix_rhotilde} is the only relevant equation for the cascade condition.

  The first-order cascade condition is given as
 \begin{align}
 (1-\mu)(1-\rho_0)\sum_{k=1}^{\lfloor 1/\theta_\lc\rfloor} \frac{kp_k}{z}(k-1) > 1, \label{eq:first-order_condition}
 \end{align}
 The irrelevancy of $F^{\agg}$ near the origin also allows us to borrow the extended cascade condition proposed by Gleeson and Cahalane~\cite{Gleeson2007}. The extended cascade condition is given by $b^2-4ac<0$, where $a, b$ and $c$ are the coefficients in the second-order approximation of Eq.(\ref{eq:appendix_rhotilde}) which is of the form $a\tilde{q}^2 +b\tilde{q}+c=0$.
 In the presence of parameter $\mu>0$, the extended cascade condition leads to (to first order in $\rho_0$)
\begin{align}\label{eq:second-order_condition}
& [(1-2\rho_0)C_1^2-4C_2(\rho_0+C_0-2C_0\rho_0)](1-\mu)^2 \nonumber \\
 & -2C_1(1-\rho_0)(1-\mu)+1 <0,
\end{align} 
 or Eq. (\ref{eq:first-order_condition}), where 
\begin{align}\label{eq:second-order_where}
 &\sum_{k=1}^{\infty}\frac{kp_k}{z}\sum_{m=0}^{k-1}B_m^{k-1}(\tilde{\rho}) F^{\lc}(m,k) \notag\\
 =& \; C_0+C_1\tilde{\rho} + C_2\tilde{\rho}^2, 
 \end{align}
 and
\begin{align}
 C_l &= \sum_{k=l+1}^\infty\sum_{m=0}^{l}\left(\!\!\begin{array}{c} k-\!1 \\ l \end{array}\!\!\right)\!
 \left(\!\begin{array}{c} l\\ m \end{array}\!\right)(-1)^{l+m}\frac{kp_k}{z}F^{\lc}(m,k).
\end{align} 
 Fig.\ref{fig:fixedthreshold_cascadecondition} illustrates the cascade region indicated by the extended cascade condition.


\section*{References}

\bibliographystyle{unsrturl}


\begin{thebibliography}{10}

\bibitem{Watts2002}
D.~J. Watts.
\newblock A simple model of global cascades on random networks.
\newblock {\em Proc. Natl. Acad. Sci. USA}, 99(9):5766--5771, 2002.
\newblock URL: \url{http://www.pnas.org/content/99/9/5766.abstract}, \href
  {http://dx.doi.org/10.1073/pnas.082090499}
  {\path{doi:10.1073/pnas.082090499}}.

\bibitem{Gleeson2007}
J.~P. Gleeson and D.~J. Cahalane.
\newblock {Seed size strongly affects cascades on random networks}.
\newblock {\em Phys. Rev. E}, 75(5):56103, 2007.
\newblock \href {http://dx.doi.org/10.1103/PhysRevE.75.056103}
  {\path{doi:10.1103/PhysRevE.75.056103}}.

\bibitem{Gleeson2008}
J.~P. Gleeson.
\newblock {Cascades on correlated and modular random networks}.
\newblock {\em Phys. Rev. E}, 77(4):46117, 2008.
\newblock \href {http://dx.doi.org/10.1103/PhysRevE.77.046117}
  {\path{doi:10.1103/PhysRevE.77.046117}}.

\bibitem{Melnik2013}
S.~Melnik, J.~A. Ward, J.~P. Gleeson, and M.~A. Porter.
\newblock {Multi-stage complex contagions}.
\newblock {\em Chaos}, 23(1):013124, 2013.
\newblock \href {http://dx.doi.org/10.1063/1.4790836}
  {\path{doi:10.1063/1.4790836}}.

\bibitem{Nematzadeh2014}
A.~Nematzadeh, E.~Ferrara, A.~Flammini, and Y.-Y. Ahn.
\newblock Optimal network modularity for information diffusion.
\newblock {\em Phys. Rev. Lett.}, 113:088701, Aug 2014.
\newblock URL: \url{http://link.aps.org/doi/10.1103/PhysRevLett.113.088701},
  \href {http://dx.doi.org/10.1103/PhysRevLett.113.088701}
  {\path{doi:10.1103/PhysRevLett.113.088701}}.

\bibitem{Liu2012}
R.~R. Liu, W.~X. Wang, Y.~C. Lai, and B.~H. Wang.
\newblock {Cascading dynamics on random networks: Crossover in phase
  transition}.
\newblock {\em Phys. Rev. E}, (85):026110, 2012.
\newblock \href {http://dx.doi.org/10.1103/PhysRevE.85.026110}
  {\path{doi:10.1103/PhysRevE.85.026110}}.

\bibitem{Payne2009}
J.~Payne, P.~Dodds, and M.~Eppstein.
\newblock {Information cascades on degree-correlated random networks}.
\newblock {\em Phys. Rev. E}, 80(2):026125, August 2009.
\newblock \href {http://dx.doi.org/10.1103/PhysRevE.80.026125}
  {\path{doi:10.1103/PhysRevE.80.026125}}.

\bibitem{Payne2011}
J.~Payne, K.~Harris, and P.~Dodds.
\newblock {Exact solutions for social and biological contagion models on mixed
  directed and undirected, degree-correlated random networks}.
\newblock {\em Phys. Rev. E}, 84(1):016110, July 2011.
\newblock \href {http://dx.doi.org/10.1103/PhysRevE.84.016110}
  {\path{doi:10.1103/PhysRevE.84.016110}}.

\bibitem{Centola2007}
D.~Centola, V.~Egu{\'\i}luz, and M.~W. Macy.
\newblock {Cascade dynamics of complex propagation}.
\newblock {\em Physica A}, 374(1):449--456, January 2007.
\newblock \href {http://dx.doi.org/doi:10.1016/j.physa.2006.06.018}
  {\path{doi:doi:10.1016/j.physa.2006.06.018}}.

\bibitem{Watts2007}
D.~J. Watts and P.~S. Dodds.
\newblock {Influentials, networks, and public opinion formation}.
\newblock {\em J. Consum. Res.}, 34(4):441--458, 2007.
\newblock \href {http://dx.doi.org/10.1086/518527} {\path{doi:10.1086/518527}}.

\bibitem{GaiKapadia2010}
P.~Gai and S.~Kapadia.
\newblock {Contagion in financial networks}.
\newblock {\em Proc. Roy. Soc. A: Math. Phys. Eng. Sci.}, 466:2401--2423, June
  2010.
\newblock \href {http://dx.doi.org/10.1098/rspa.2009.0410}
  {\path{doi:10.1098/rspa.2009.0410}}.

\bibitem{Gai2011}
P.~Gai, A.~Haldane, and S.~Kapadia.
\newblock {Complexity, concentration and contagion}.
\newblock {\em J. Monetary Econ.}, 58:453--470, July 2011.
\newblock \href {http://dx.doi.org/10.1016/j.jmoneco.2011.05.005}
  {\path{doi:10.1016/j.jmoneco.2011.05.005}}.

\bibitem{Kobayashi2014}
T.~Kobayashi.
\newblock {A model of financial contagion with variable asset returns may be
  replaced with a simple threshold model of cascades}.
\newblock {\em Econ. Lett.}, 124:113--116, 2014.
\newblock \href {http://dx.doi.org/10.1016/j.econlet.2014.05.003}
  {\path{doi:10.1016/j.econlet.2014.05.003}}.

\bibitem{Brummitt2015}
C.~D. Brummitt and T.~Kobayashi.
\newblock Cascades in multiplex financial networks with debts of different
  seniority.
\newblock {\em Phys. Rev. E}, 91:062813, Jun 2015.
\newblock URL: \url{http://link.aps.org/doi/10.1103/PhysRevE.91.062813}, \href
  {http://dx.doi.org/10.1103/PhysRevE.91.062813}
  {\path{doi:10.1103/PhysRevE.91.062813}}.

\bibitem{Lee2014}
K.-M. Lee, C.~D. Brummitt, and K.-I. Goh.
\newblock {Threshold cascades with response heterogeneity in multiplex
  networks}.
\newblock {\em Phys. Rev. E}, 90(6):062816, December 2014.
\newblock \href {http://dx.doi.org/10.1103/PhysRevE.90.062816}
  {\path{doi:10.1103/PhysRevE.90.062816}}.

\bibitem{Yagan2012}
O.~Ya\u{g}an and V.~Gligor.
\newblock {Analysis of complex contagions in random multiplex networks}.
\newblock {\em Phys. Rev. E}, 86(3):036103, September 2012.
\newblock \href {http://dx.doi.org/10.1103/PhysRevE.86.036103}
  {\path{doi:10.1103/PhysRevE.86.036103}}.

\bibitem{Brummitt2012_PRER}
C.~D. Brummitt, K.-M. Lee, and K.-I. Goh.
\newblock {Multiplexity-facilitated cascades in networks}.
\newblock {\em Phys. Rev. E}, 85:045102(R), April 2012.
\newblock \href {http://dx.doi.org/10.1103/PhysRevE.85.045102}
  {\path{doi:10.1103/PhysRevE.85.045102}}.

\bibitem{Strogatz2014}
S.~H. Strogatz.
\newblock {\em Nonlinear dynamics and chaos: with applications to physics,
  biology, chemistry, and engineering}.
\newblock Westview press, 2014.

\bibitem{Riedel1972}
E.~K. Riedel and F.~J. Wegner.
\newblock Tricritical exponents and scaling fields.
\newblock {\em Phys. Rev. Lett.}, 29:349--352, Aug 1972.
\newblock URL: \url{http://link.aps.org/doi/10.1103/PhysRevLett.29.349}, \href
  {http://dx.doi.org/10.1103/PhysRevLett.29.349}
  {\path{doi:10.1103/PhysRevLett.29.349}}.

\bibitem{Araujo2011}
N.~A.~M. Ara\'ujo, J.~S. Andrade, R.~M. Ziff, and H.~J. Herrmann.
\newblock Tricritical point in explosive percolation.
\newblock {\em Phys. Rev. Lett.}, 106:095703, Mar 2011.
\newblock URL: \url{http://link.aps.org/doi/10.1103/PhysRevLett.106.095703},
  \href {http://dx.doi.org/10.1103/PhysRevLett.106.095703}
  {\path{doi:10.1103/PhysRevLett.106.095703}}.

\bibitem{Cellai2011}
D.~Cellai, A.~Lawlor, K.~A. Dawson, and J.~P. Gleeson.
\newblock Tricritical point in heterogeneous $k$-core percolation.
\newblock {\em Phys. Rev. Lett.}, 107:175703, Oct 2011.
\newblock URL: \url{http://link.aps.org/doi/10.1103/PhysRevLett.107.175703},
  \href {http://dx.doi.org/10.1103/PhysRevLett.107.175703}
  {\path{doi:10.1103/PhysRevLett.107.175703}}.

\end{thebibliography}


\end{document}